\documentclass[twocolumn,iop,revtex4]{openjournal}
\usepackage[utf8]{inputenc}
\usepackage{times}
\usepackage{natbib} 
\usepackage{epsfig} 
\usepackage{graphicx} 
\usepackage[dvipsnames]{xcolor}
\usepackage{aas_macros} 
\usepackage{amssymb}
\usepackage{amsmath}
\usepackage[title]{appendix}
\usepackage{hyperref}	
\hypersetup{colorlinks=true,linkcolor=blue,citecolor=blue,filecolor=blue,urlcolor=blue}
\usepackage[caption=false]{subfig}
\usepackage{soul}                          
\usepackage{ulem} 
\usepackage{xifthen}

\newcommand{\enzo}{\texttt{Enzo~}}
\newcommand{\yt}{\texttt{yt}}

\newcommand{\msolar} {$\rm{M_{\odot}}~$}
\newcommand{\msolarc} {$\rm{M_{\odot}}$}
\newcommand{\msolaryr} {$\rm{M_{\odot}~yr^{-1}}~$}
\newcommand{\msolaryrc} {$\rm{M_{\odot}~yr^{-1}}$}

\newcommand{\zsolar} {$\rm{Z_{\odot}}~$}
\newcommand{\zsolarc} {$\rm{Z_{\odot}}$}

\newcommand{\molH} {$\rm{H_2}$~}

\newcommand{\rarepeak} {\textit{Rarepeak~}}
\newcommand{\normal} {\textit{Normal~}}
\newcommand{\void} {\textit{Void~}}
\newcommand{\voidc} {\textit{Void}}

\newcommand{\change}[2][]{%
\ifthenelse{\isempty{#2}}{{\color{ForestGreen}{#1}}}%
{{\color{RedOrange}\sout{#1}}{\color{ForestGreen}{#2}}}%
}

\begin{document}
\title{Massive Star Formation in Metal-Enriched Haloes at High Redshift}
\author{John A. Regan$^{1, *}$}
\thanks{$^*$E-mail:john.regan@mu.ie, Royal Society - SFI University Research Fellow}
\author{Zolt\'an Haiman$^{2}$}
\author{John H. Wise$^{3}$}
\author{Brian W. O'Shea$^{4,5,6,7}$}
\author{Michael L. Norman$^8$}

\affiliation{$^1$Department of Theoretical Physics, Maynooth University, Maynooth, Ireland}
\affiliation{$^2$Department of Astronomy, Columbia University, 550 W. 120th Street, New York, NY 10027, USA}
\affiliation{$^3$Center for Relativistic Astrophysics, Georgia Institute of Technology, 837 State Street, Atlanta, GA 30332, USA}
\affiliation{$^4$Department of Computational Mathematics, Science, and Engineering, Michigan State University, MI, 48823, USA}
\affiliation{$^5$Department of Physics and Astronomy, Michigan State University,MI, 48823, USA}
\affiliation{$^6$Joint Institute for Nuclear Astrophysics - Center for the Evolution of the Elements, USA}
\affiliation{$^7$National Superconducting Cyclotron Laboratory, Michigan State, University, MI, 48823, USA}
\affiliation{$^8$Center for Astrophysics and Space Sciences, University of California, San Diego, 9500 Gilman Dr, La Jolla, CA 92093}

\begin{abstract}
  The formation of supermassive stars has generally been studied under the assumption of 
  rapid accretion of pristine metal-free gas. Recently it was found, however, that gas enriched
  to metallicities up to $Z \sim 10^{-3}$ \zsolar can also facilitate supermassive star formation,
  as long as the total mass infall rate onto the protostar remains sufficiently high. We extend the
  analysis further by examining how the abundance of supermassive star candidate haloes would be
  affected if all haloes with super-critical infall rates, regardless of metallicity
  were included. We investigate this scenario by identifying all atomic cooling haloes in the
  Renaissance simulations with central mass infall rates exceeding a fixed threshold.
  We find that among these haloes with central mass infall rates above 0.1 \msolaryrc\ approximately
  two-thirds of these haloes have metallicities of $Z > 10^{-3}$ \zsolarc.
  If metal mixing within these haloes is
  inefficient early in their assembly and pockets of metal-poor gas can remain then the
  number of haloes hosting supermassive stars can be increased by at least a factor of four.
  Additionally the centres of these high infall-rate haloes provide ideal environments in which to grow
  pre-existing black holes. Further research into the (supermassive) star
  formation dynamics of rapidly collapsing haloes, with inhomogeneous
  metal distributions, is required to gain more insight into both supermassive star
  formation in early galaxies as well as early black hole growth.   
\end{abstract}

\section{Introduction} \label{Sec:Introduction}
\noindent Supermassive black holes (SMBHs) with masses exceeding $10^{9}$~\msolar have long been known to
exist at high redshift \citep[e.g.][]{Fan_2001, Dietrich_2002, Fan_2003, Vestergaard_2004,
  Fan_2004, Fan_06}. The number of quasars, powered by SMBHs, at $z > 6$ now exceeds
200 \citep{Matsuoka_2019}\footnote{For recent comprehensive compilations, see \citet{Inayoshi_2020} and
\url{http://www.sarahbosman.co.uk/list_of_all_quasars}.}
and their number
density is estimated to be of order 1~Gpc$^{-3}$. These 200 or so observed quasars probably
represent the tip of the iceberg with many more quasars of lower luminosity (and hence perhaps mass)
lurking below the observational threshold of current telescopes. \\
\indent Nonetheless, the discovery of even this relatively small number of high redshift SMBHs
has posed a significant challenge to our understanding of the formation and growth of black holes.
Black holes are believed to form as the endpoint of massive stars. The seed black hole may then grow through the
accretion of gas or through mergers with other black holes. The significant challenge in this
scenario is that SMBHs with masses of up to $10^{9}$~\msolar exist \citep[e.g][]{Banados_2018} already when
the Universe was less than one billion years old. Our current understanding of black hole accretion
therefore makes these observations difficult to interpret. Three mainstream scenarios have
emerged over the past two decades in an attempt to understand the origin of high-z SMBHs (see \citealt{Inayoshi_2020} for a recent extended review). \\
\indent Perhaps the most straightforward scenario is that all black holes emerge from the end
point of the first generation of stars. The initial mass function (IMF) of the first stars
is currently unknown, and as yet we have little direct observational guidance. However, simulations,
as well as the lack of any observed zero-metallicity stars in our Galaxy, point to an IMF that
is top heavy compared to present-day star formation \citep{Yoshida_2006, Turk_2009,
  Clark_2011a, Hirano_2014}. While initial studies of Population III (PopIII) star formation
concluded that the masses of the first stars were of the order of 100 \msolar \citep{Bromm_1999,
  Abel_2002, Bromm_2002} more recent studies indicate a characteristic mass of a few tens of solar
masses but with a range into the hundreds of solar masses \citep{Stacy_2010, Stacy_2012, Stacy_2014,
  Hirano_2014,Hirano_2015}. A black hole born from such stars will have at most the same mass as the progenitor star and hence 
limited to
a few tens to a few hundred solar masses. These black holes are known as ``light seeds'' \citep[e.g.][]{Volonteri_2010a},
and will need to increase their mass by up to
eight orders of magnitude within a few hundred million years if they are to grow to become a SMBH. Such
a scenario has been shown to require substantial fine-tuning~\citep{Tanaka_2009,Tanaka_2012}.
Black holes born from PopIII remnants exist
in relic ionized regions that are warm and diffuse, inhospitable for growth
\citep{Whalen_2004, Milosavljevic_2009, Alvarez_2009}. Furthermore, after these seeds are incorporated into galaxies,
they often find themselves in underdense regions away from the halo centre \citep{Smith_2018}.
The black holes
are unable to grow efficiently and suffer accretion rates orders of magnitudes below the Eddington
rate until its host halo is replenished with gas. 
Likewise, growing these black holes via mergers is difficult, due to the gravitational-wave recoil
which tends to eject the merger remnant from the shallow potential of their host halo~\citep{Haiman_2004}.
It would seem that only in the cases where these 
black holes can grow at rates significantly
exceeding the Eddington limit can these black holes reach sufficiently high masses. While extended episodes of modestly 
super-Eddington accretion \citep{Madau_2001, Madau_2014, Alexander_2014,
  Lupi_2016}, as well as shorter episodes of hyper-Eddington accretion~\citep{Inayoshi_2018}
remain physically viable, it is unclear how often they are realised in nature with a
sufficient duty cycle~\citep{Pacucci_2017}.

\indent Scenarios two and three are the so-called ``heavy seed" scenarios, both of which create black holes with initial masses of
greater than 1000 \msolar and possibly up to $10^{6}$ \msolarc. Scenario two is the collapse of a dense
nuclear star cluster \citep{PortegiesZwart_2004, Freitag_2008, Omukai_2008,Devecchi_2008, Merritt_2008,
  Davies_2011, Lupi_2014}. In this case the compact nature of the cluster allows
stellar collisions to dominate resulting in the formation of a very massive seed at the
centre of the halo. Detailed numerical simulations have been conducted \cite[e.g.][]{Katz_2015,
  Reinoso_2018, Sakurai_2019} of this scenario and
most studies have converged on a final mass of the central object of approximately 1000 \msolarc, constituting a small fraction of the mass of the gas in an atomic-cooling halo.
Whether this scenario can ultimately produce SMBHs at early times in the Universe is as of yet
unclear and further research of this scenario is still required. \\
\begin{figure*}
\centering
\begin{minipage}{175mm}      \begin{center} 
\centerline{
\includegraphics[width=0.525\textwidth]{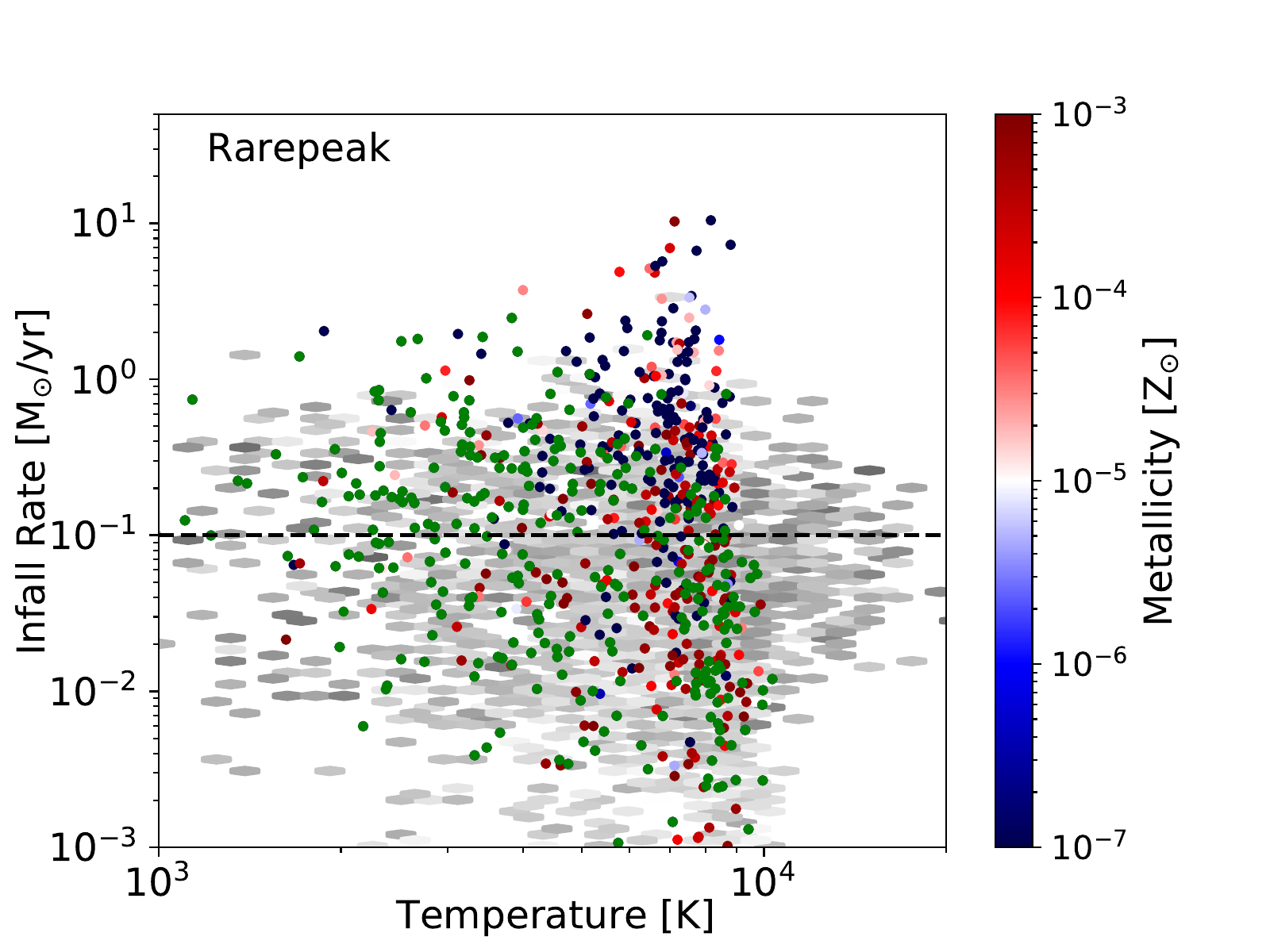}
\includegraphics[width=0.525\textwidth]{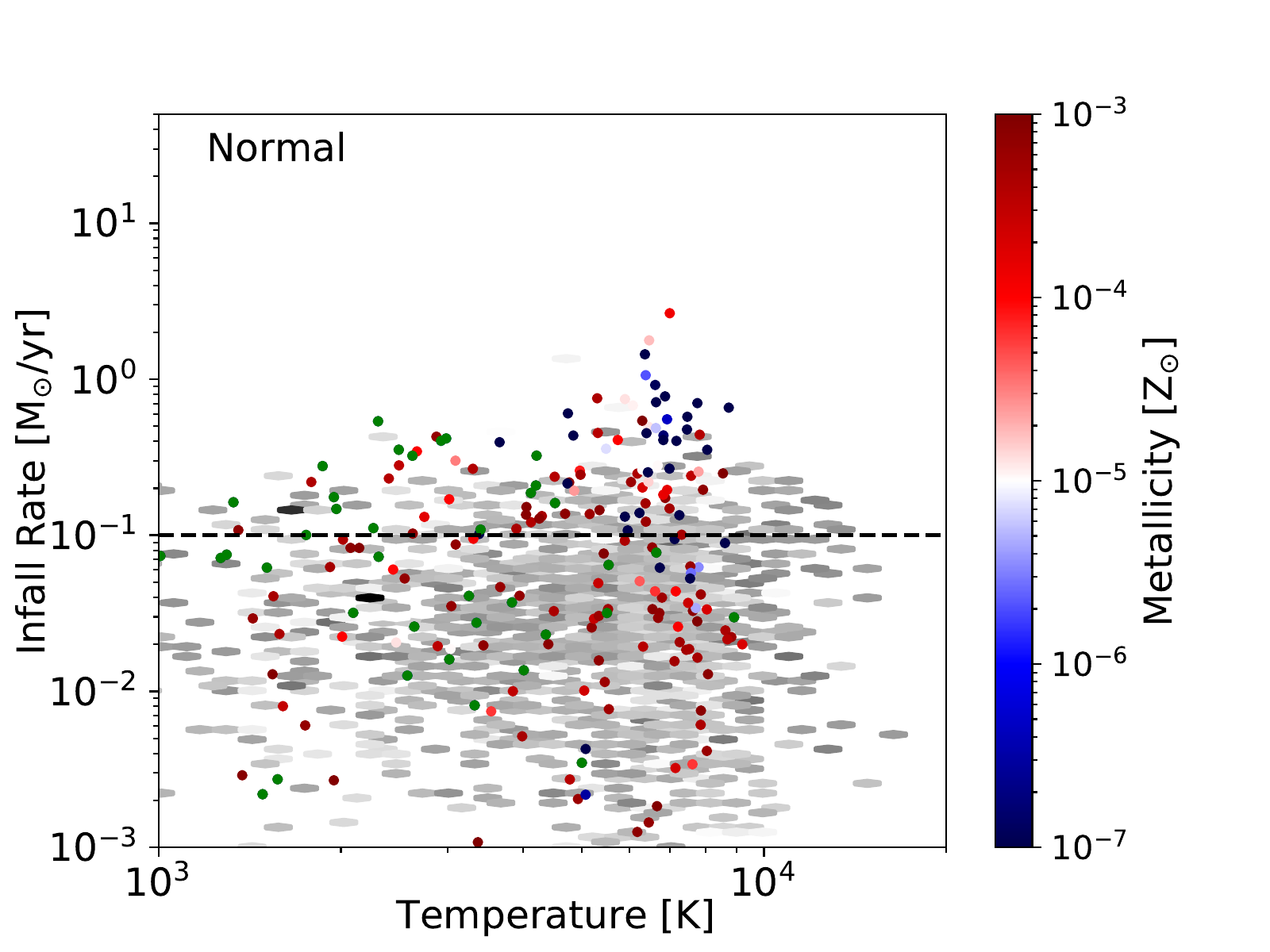}}
\caption{\textit{Left Panel}: The mass infall rate inside 20 pc for \textit{all} 
  3244 halos between $15<z<24$ in 
  the \rarepeak simulation. The infall rates are plotted
  against the volume averaged temperature of the gas within 20 pc, colour coded by metallicity. Hexbins, in
  grayscale, are used to represent the metal-enriched haloes ($Z > 10^{-3}$ \zsolarc).
  Darker colours signify higher metallicity as well as higher halo counts. Blue, red and green circles
  represent the lower metallicity haloes. Red circles represent haloes metallicities
  with $10^{-3} $ \zsolar $\gtrsim Z \gtrsim 10^{-5}$ \zsolar while blue circles represent
  haloes with $Z \lesssim 10^{-5}$ \zsolarc. Green circles represent haloes which are both
  metal-free \underline{and} star-free.  \textit{Right Panel}: The same plot for the 2335 halos between $11.6<z<24$ in the \normal simulation
  for the infall rate and temperatures within 20 pc of the halo centres.
  The \normal simulation has fewer very high ($ \gtrsim 1 $ \msolaryrc) mass infall rate haloes.
  The majority of haloes in the \normal region are also metal-enriched, with fewer extremely metal-poor and
  metal-free haloes by number (but not by percentage - see Figure \ref{Fig:Fractions}) compared to the \rarepeak region.
  The black dashed line in each plot marks the 0.1 \msolaryr threshold value. This marks a distinction
  between high infall rate haloes and lower infall rate haloes.}
  \label{Fig:Scatter}
\end{center} \end{minipage}

\end{figure*}
\indent A third scenario, proposed to create yet heavier seeds, is arguably the most well studied.
In this scenario very high accretion rates onto embryonic protostars allows for the creation of supermassive stars (SMSs)
\citep{Shapiro_1979, Begelman_2008, Schleicher_2013, Hosokawa_2013, Inayoshi_2014, Sakurai_2016,
  Umeda_2016, Haemmerle_2018,Woods_2017, Woods_2018, Regan_2018b}. \cite{Haemmerle_2018} recently found
that a mean accretion rate above a critical value of 0.001 \msolaryr is required to create a SMS.
If this accretion is maintained, the stellar atmosphere inflates and the star is bloated,
causing the surface temperatures to drop to approximately 5000 K \citep{Hosokawa_2013, Woods_2017, Haemmerle_2018},
thus avoiding the deleterious feedback effects from ionizing radiation.\\
\indent In order to achieve the large required accretion rates, the general assumption has been
that a metal-free atomic cooling halo is required, whose gas remains near its virial temperature of $\sim10^4$K.
With the mass accretion rate onto the central regions of a halo scaling as $dM/dt \propto T^{3/2}$ \citep{Shu_1977}, such haloes would appear
to be optimal regions to support rapid mass infall. The absence of metals removes a cooling pathway for
the gas which would otherwise potentially cause the gas to cool and fragment into smaller, lower
mass, stars thus preventing the formation of a massive central object. A further criteria is that the
atomic cooling halo should have had no previous episode of star formation (which would have
internally enriched the halo with metals). In order to maintain the gas temperature near the virial temperature of such a pristine halo, the \molH content
must also be suppressed. Many studies have investigated pathways to achieve this, the main pathways
investigated have been to employ strong Lyman-Werner (LW) fluxes \citep{Dijkstra_2008, Shang_2010,
  Regan_2014b, Latif_2014b, Agarwal_2015a, Latif_2015, Regan_2016a, Regan_2017, Regan_2018a},
baryonic streaming velocities
\citep{Tseliakhovich_2010, Tanaka_2014, Hirano_2017, Schauer_2017}, collisions of massive
proto-galaxies \citep{Mayer_2010, Mayer_2014, Inayoshi_2015} and finally dynamical heating caused
by a succession of minor and major mergers \citep{Yoshida_2003a, Fernandez_2014, Wise_2019}. \\
\indent The goal of the present study is to examine the role that metal enrichment in embryonic
haloes may play in the
formation of SMSs. We adopt similar terminology from the existing literature
on metal-poor stars \citep[e.g.][]{Frebel_2015}. Our definitions differ slightly from those of
\cite{Frebel_2015} - particularly by including a ``metal-free'' category which is generally
absent when considering stellar metallicities. We assign the following terms to halo metallicities:
\begin{itemize}
\item metal-free $\equiv Z < 10^{-5}$ \zsolarc 
\item extremely metal-poor (EMP) $\equiv 10^{-5}$ \zsolar $< Z < 10^{-3}$ \zsolarc 
\item metal-enriched\footnote{The metal-poor stars community traditionally includes stars with
  metallicities of $Z > 10^{-3}$ \zsolar in the metal-poor category, but we refer to these stars as metal-enriched, which 
  is more appropriate in the present context.} $\equiv Z > 10^{-3}$ \zsolarc.
\end{itemize}
  The formation of SMSs in haloes with metallicities in the EMP range
  has recently been investigated by \cite{Tagawa_2020} and \cite{Chon_2020}. \cite{Tagawa_2020}
  utilised semi-analytic models, and showed that if the gas fragments at sufficiently high
  density ($\gtrsim {\rm few}\times 10^{10}~{\rm cm^{-3}}$), the central protostar avoids contraction
  and grows via frequent mergers with other protostars into a $\sim 10^5-10^6~{\rm M_\odot}$ SMS.
  This represents a hybrid scenario between the collapse of a dense stellar cluster and the
  growth of a single SMS discussed above. \cite{Chon_2020} used high-resolution smoothed particle
  hydrodynamics simulations to investigate a similar scenario. They followed the cooling,
  contraction, and fragmentation of gas in the high-density core of a massive halo with a high
  gas inflow rate ($dM/dt=1$~\msolaryrc). \cite{Chon_2020} investigate the effects of
    both metallicity and dust in their simulations. The impact of
    dust cooling is investigated using the prescription from \cite{Omukai_2008}, which operates
    at gas number densities of $\rm{n_{gas} \gtrsim 10^{10} \ cm^{-3}}$. They find
    that the impact of dust cooling does not extend to scales larger than a few hundred
    au\footnote{Dust cooling is not included in the Renaissance simulations as we do not reach densities
      high enough to activate that cooling channel}. Their overall findings are that the formation
  pathway of massive stars depends strongly on metallicity: the growth of the SMS is dominated by
  gas accretion in the metal-free case, while mergers with other fragments become dominant as
  the metallicity approaches the upper limit of the EMP range. They found that SMS formation was
  prevented at the highest metallicity they studied, $Z=10^{-3}$ \zsolarc. However, at all
  lower metallicities, the end result is the same: the rapid growth of a massive central protostar,
  at an average rate of $\sim$1 \msolaryrc.\\
  \indent Overall, the above suggests that {\em SMS formation in the dense core of an atomic halo is possible,
    even when it has significant metallicity}, as long as the infall rate is high and fragmentation occurs at sufficiently high density.
This raises the question: {\em how many additional such haloes there are in the early universe, compared to metal-free haloes?}
To assess the abundance and demography of such haloes at high redshift, we examine the outputs of the Renaissance Simulations
and look for haloes that display rapid collapse, regardless of the metallicity of the halo.
We examine rapidly accreting haloes across all redshifts and record their metallicities and other properties.
Our analysis expands on similar earlier searches for supermassive star candidate haloes
(e.g.,~\citealt{Habouzit_2016,Tremmel_2017,Dunn_2018,Wise_2019,Regan_2020}) into the regime of haloes that are enriched with
metals, selecting candidates based on their infall rates, rather than on the requirement of being metal-free.

This paper is organised as follows.
In \S~\ref{Sec:RenaissanceDatasets}, we describe the datasets that comprise the Renaissance Simulations
and our analysis methods.
In \S~\ref{Sec:Results}, we present  our main results, in the form of the demography of
haloes with rapid central gas inflow rates, and their distribution in metallicity.  
Finally, in \S~\ref{Sec:Discussion}, 
we discuss our findings further and summarise the main conclusions and implications of this work.

\section{Renaissance Datasets} \label{Sec:RenaissanceDatasets}
The Renaissance Simulations were carried out on the Blue Waters supercomputer facility using the
adaptive mesh refinement code \enzo\citep{Enzo_2014, Enzo_2019}\footnote{\url{https://enzo-project.org/}}.
\enzo has been extensively used to study the formation of structure in the early universe
\citep{Abel_2002, OShea_2005b, Turk_2012, Wise_2012b, Wise_2014, Regan_2015, Regan_2017}. In
particular, \enzo includes a ray-tracing scheme to follow the propagation of radiation from star
formation and
black hole formation \citep{WiseAbel_2011} as well as a detailed multi-species chemistry model that
tracks the formation and evolution of nine species \citep{Anninos_1997, Abel_1997, Grackle}.
Additionally the photo-dissociation of \molH is followed, which is a critical ingredient for
determining the formation of the first metal-free stars \citep{Haiman_2000}.

\begin{figure*}
\centering
\begin{minipage}{175mm}      \begin{center} 
\centerline{
\includegraphics[width=0.525\textwidth]{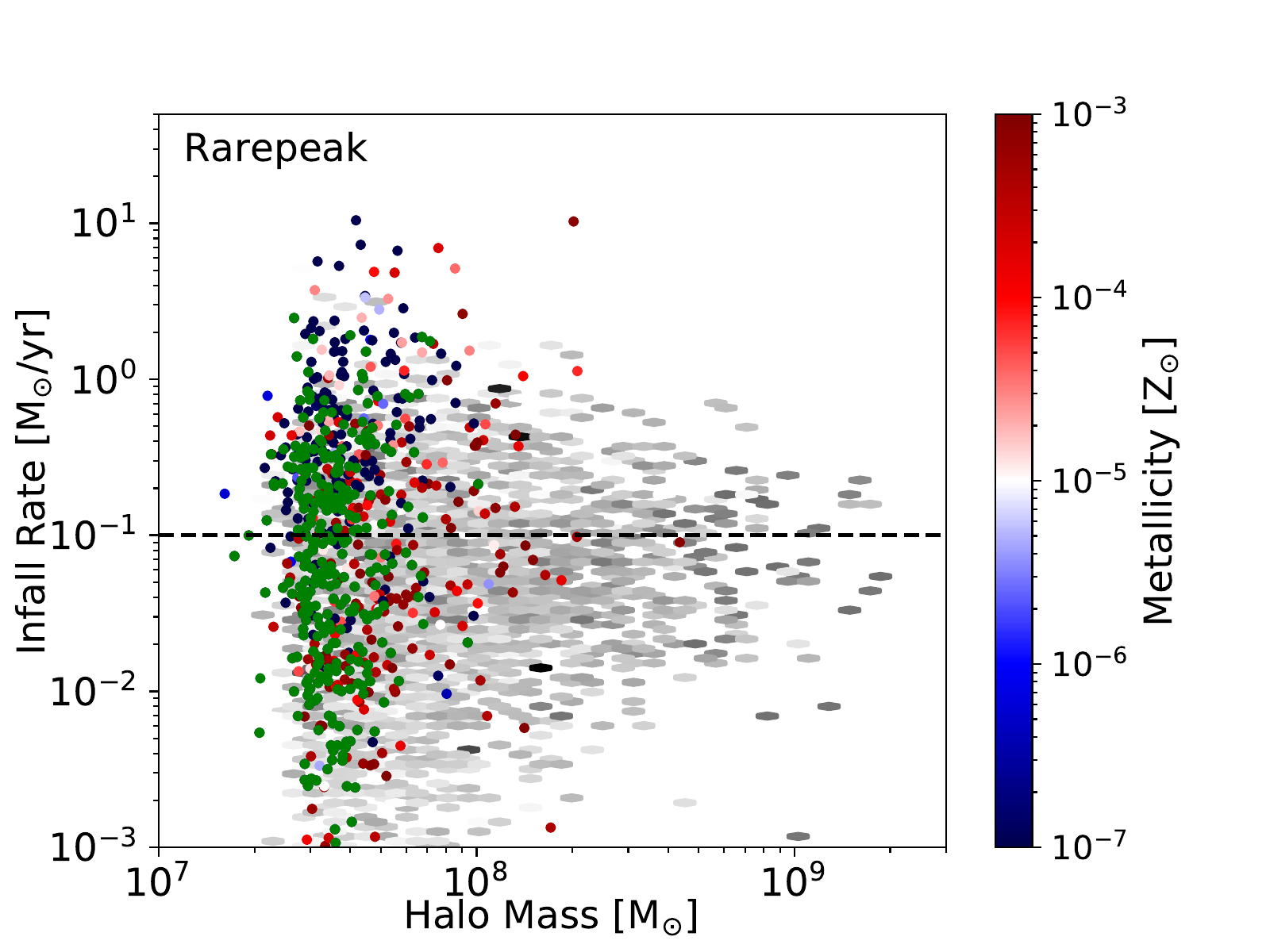}
\includegraphics[width=0.525\textwidth]{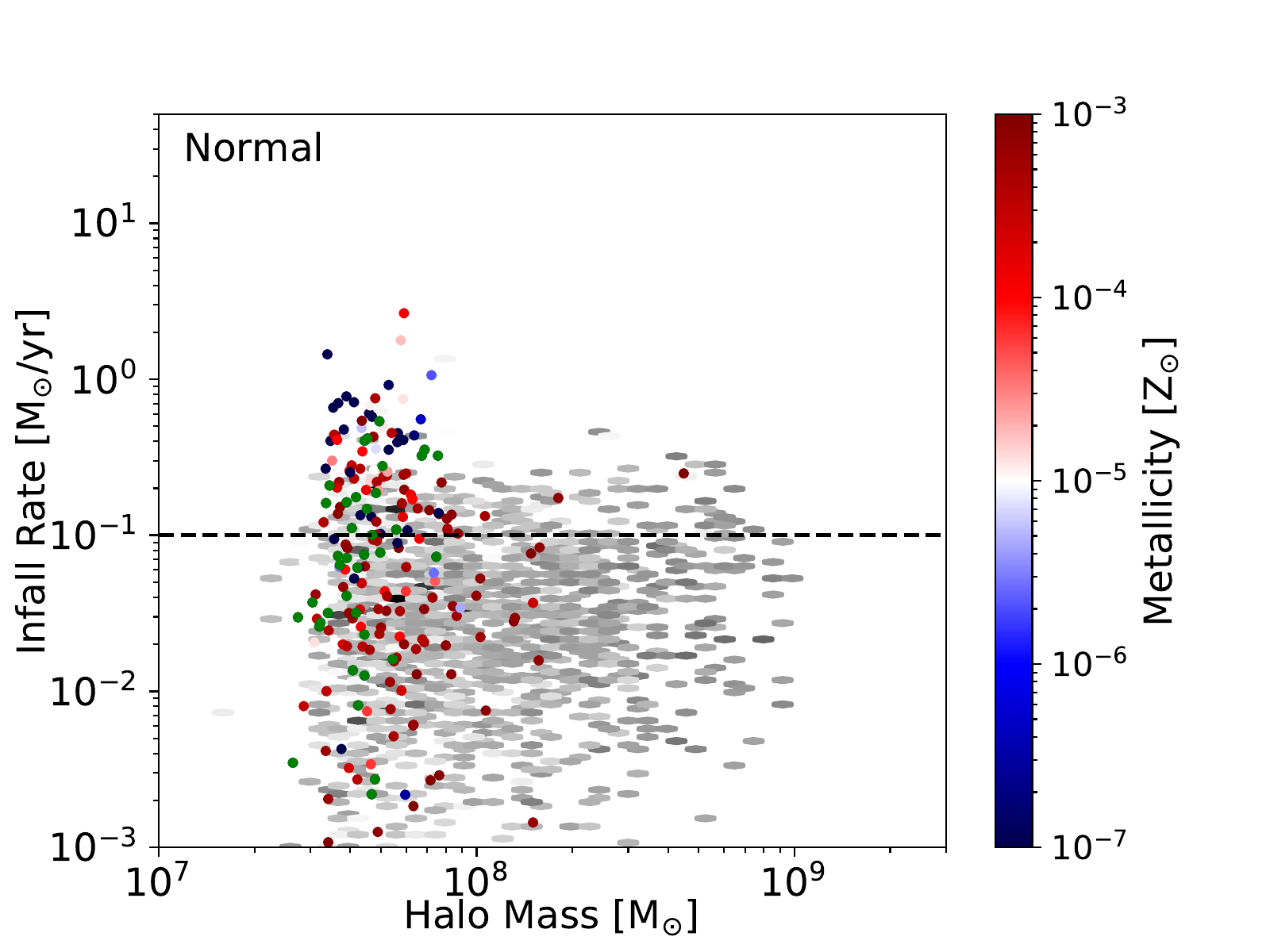}}
\caption{Same as Figure~\ref{Fig:Scatter}, except
the infall rates are plotted against the mass of each parent halo.
In the \rarepeak region (\textit{left panel}), the extremely metal-poor
  and metal-free haloes ($Z < 10^{-3}$ \zsolarc; shown in red and blue colours) populate the halo mass range predominantly just below $10^{8}$
  \msolarc.  The \normal region (\textit{right panel})
 has significantly fewer extremely metal-poor and metal-free haloes in this halo mass range but nonetheless the trend is similar.}
\label{Fig:HaloMass}
\end{center} \end{minipage}

\end{figure*}

The datasets used in this study were originally derived from a simulation of the universe in a (40 Mpc)$^3$
box using the WMAP7 best fit cosmology~\citep{Komatsu_2011}.
Initial conditions were generated using
MUSIC \citep{Hahn_2011} at redshift $z = 99$. A low-resolution simulation was run until $z = 6$ in order to
identify three different regions for re-simulation \citep{Chen_2014}. The volume was then smoothed
on a physical scale of 5 comoving Mpc, and regions of high
($\langle\delta\rangle \equiv \langle\rho\rangle/(\Omega_{\rm M} \rho_{\rm c}) - 1 \simeq 0.68$),
average ($\langle\delta\rangle \sim 0.09$), and low ($\langle\delta\rangle \simeq -0.26$)
mean overdensity were chosen for re-simulation. These sub-volumes are referred to as the
\rarepeak region, the \normal region  and the \void region. The \rarepeak region has a comoving
volume of 133.6 Mpc$^3$, and the \normal and \void regions both have comoving volumes of 220.5
Mpc$^3$. Each region was then re-simulated with an effective initial resolution of $4096^3$ grid
cells and particles within these sub-volumes of the larger initial simulation. This gives a maximum
dark matter particle mass resolution of $2.9 \times 10^4$ \msolarc. For the re-simulations, further refinement was allowed throughout the sub-volumes up
to a maximum refinement level of 12, which corresponds to 19 pc comoving spatial resolution. Given
that the
\voidc, \normal and \rarepeak regions focus on different overdensities, each region was evolved forward in time to
different epochs. The \rarepeak region, being the most overdense and hence the most
computationally demanding at earlier times, was run until $z = 15$. The \normal region ran until $z =
11.6$, and the \void region ran until $z = 8$. In all of the regions the halo mass function was 
well resolved down to M$_{\rm halo} \sim 2 \times 10^6$ \msolarc. The \rarepeak region represents a
volume which is 1.68 times denser than the cosmic mean - therefore at $z = 15$, regions with the
size and overdensity of the \rarepeak are expected to be found in approximately 0.01-–0.1\%
of the volume of the Universe \citep{Wise_2019}. Therefore, any results from the
\rarepeak region need to be viewed in the context of the region being rare ($\sim 10^3 - 10^4$ times rarer than the \normal region).
The \rarepeak regions contains 822 galaxies with masses larger than $2 \times 10^7$ \msolar
at $z = 15$, the \normal region contains 758 such galaxies at $z = 11.6$, while the \void region
contains 458 such galaxies at $z = 8$. In this study, we examine only the \rarepeak and \normal
regions. In our previous work
\citep{Wise_2019, Regan_2020} we found the \void region to be devoid of metal-free and
star-free atomic cooling haloes as infall rates
did not support their formation, we therefore do not include the \void region in this study.

\section{Results} \label{Sec:Results}

\subsection{Atomic cooling haloes in the Renaissance simulation}

As discussed in \S \ref{Sec:RenaissanceDatasets} the \rarepeak dataset represents an overdense
region of the Universe with a large selection of galaxies while the \normal region represents a
region with an average density comparable to the cosmic mean. We begin our analysis of these two
datasets by examining only those haloes which lie inside the high-resolution regions and which
have masses greater than the atomic cooling limit, which we adopt from \cite{Fernandez_2014},
\begin{equation}
  M_{\rm atm} = 2 \times 10^{7} \left( \frac{{T_{\rm vir}}}{10^4~{\rm K}}  \frac{21}{1+z} \right)^{-1.5} \rm{M_{\odot}},
\end{equation}
that is calibrated with simulations, where $z$ is the redshift and $T_{\rm vir}$ is the virial temperature of the halo. 
We tested other definitions \citep[e.g.][]{Bromm_2011} which give larger
values by up to a factor of two or more. The number of metal-free haloes
is very sensitive to the chosen value because it lies on the 
relatively rare
exponential part of the halo mass function.
We chose a value of the atomic limit consistent with previous work in the literature, and
calibrated to correspond to the onset of Lyman $\alpha$ cooling in cosmological simulations,
but we note that this choice impacts our results.
For both datasets, we then filter each halo based on the mass infall rate in that halo.
We calculate the instantaneous mass infall rate using the standard continuity equation:
\begin{equation}  
  \dot{M} = 4 \pi R^2 \rho(R) V_{\rm rad}(R)
\end{equation}
where $\dot{M}$ is the mass infall rate, $R$ is the radius at which the infall rate is calculated,
$\rho(R)$ is the gas density at that radius and $V_{\rm rad}(R)$ is the radial velocity at that radius.
For both the \rarepeak and \normal region the mass infall rate is calculated as the mean
infall rate within 20 pc of the centre of the halo. 

The value of 20 pc was dictated by the spatial resolution of our simulations, and
was chosen to give a large enough
volume over which to calculate the mean while also focusing on the region in which SMS formation is
expected. \\
\indent The \rarepeak simulation was run until $z=15$, but we
examine all of the outputs\footnote{There are 40 \rarepeak outputs between $z = 24$ and $z = 15$.}
from the dataset. For the \normal datasets we adopt the exact same procedure except that
the \normal simulation runs until $z = 11.6$ and again we examine all of the
outputs\footnote{There are 70 \normal outputs between $z = 24$ and $z = 11.6$.}.  In total, we identified
3,244 atomic cooling haloes across all outputs in the \rarepeak region and 2,335
haloes across all outputs in the \normal region. These numbers represent haloes at various stages
of their growth and accretion rate history. 
\begin{figure*}
\centering
\begin{minipage}{175mm}      \begin{center} 
\centerline{
\includegraphics[width=0.525\textwidth]{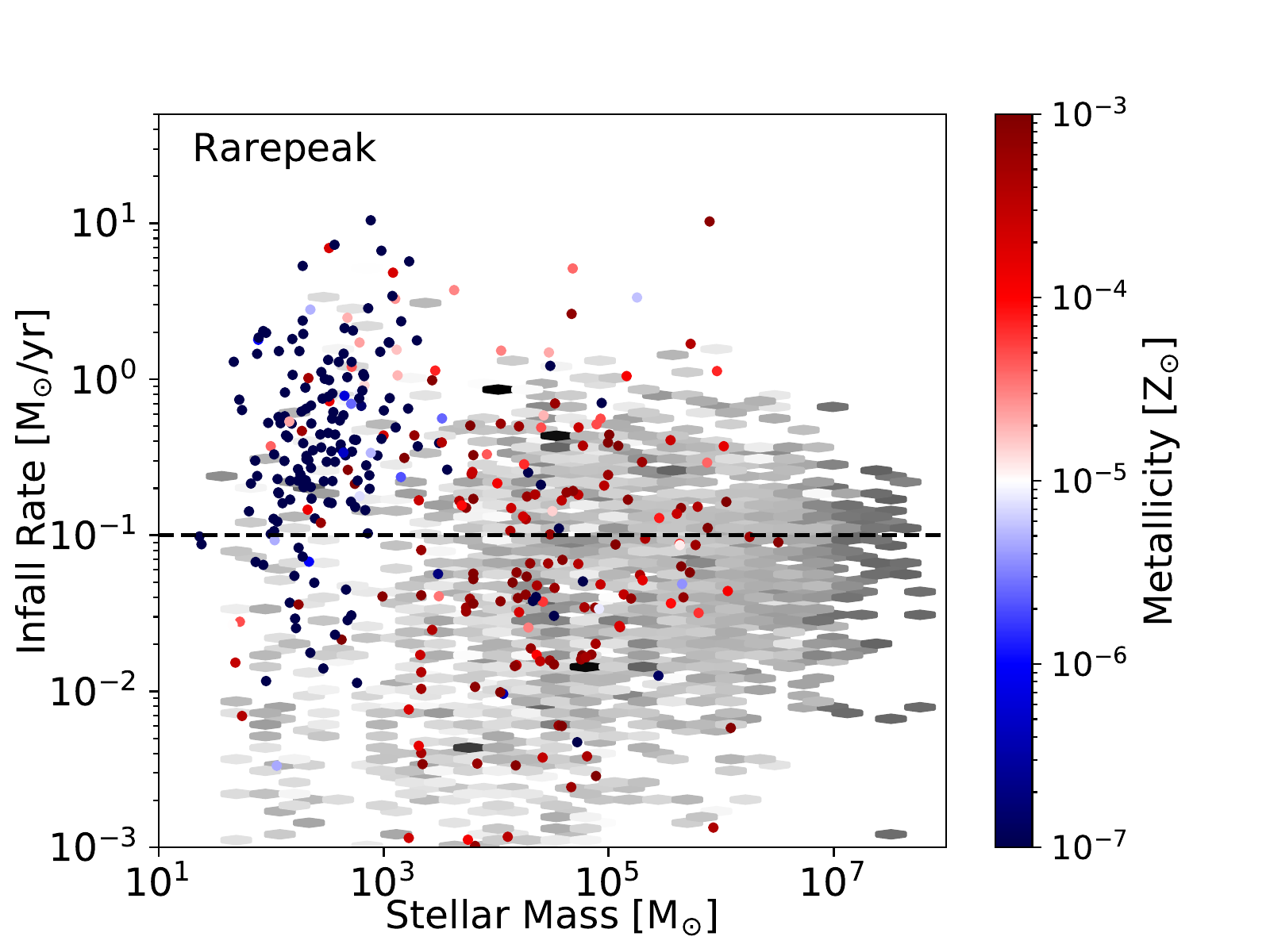}
\includegraphics[width=0.525\textwidth]{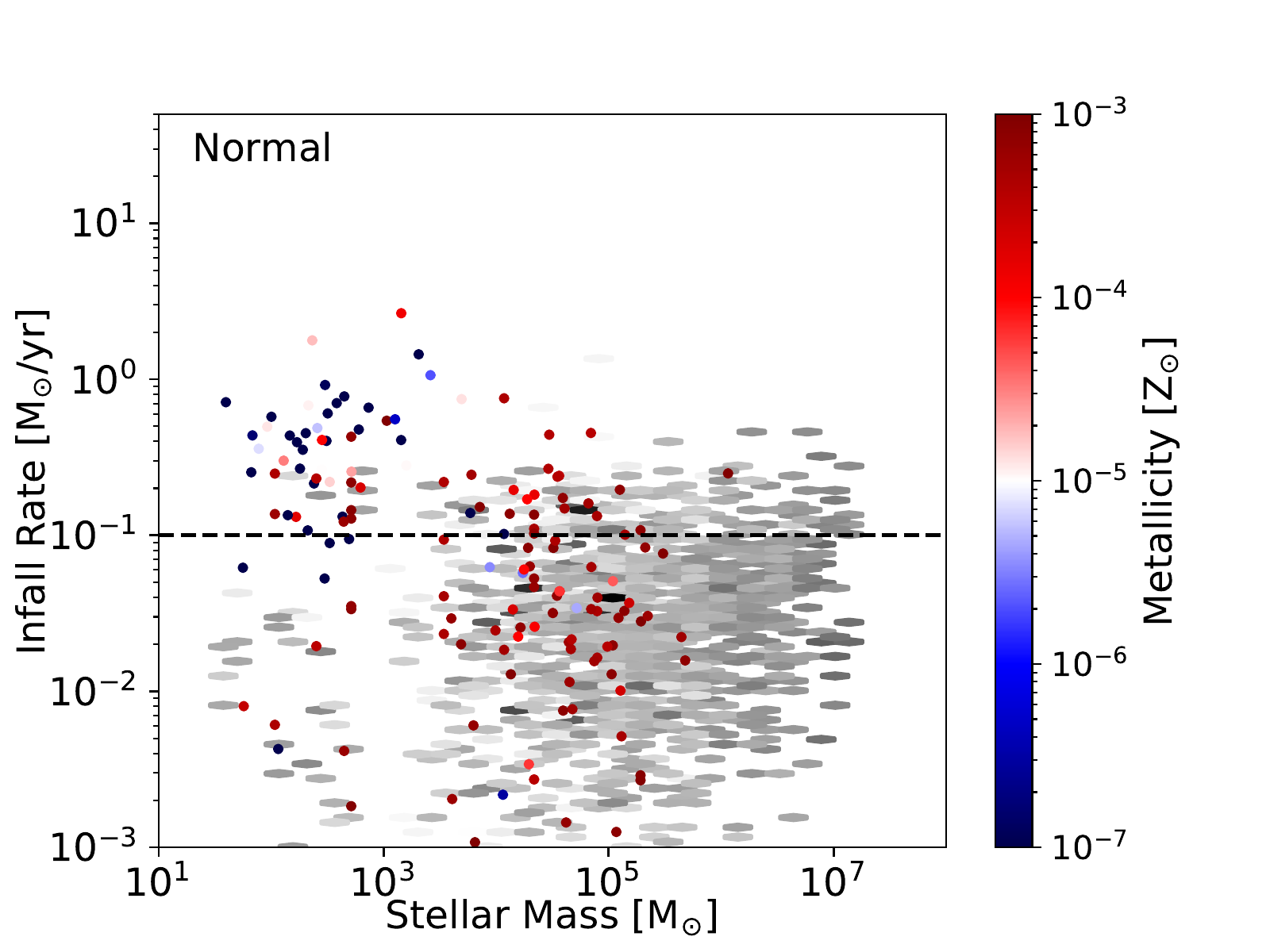}}
\caption{Same as Figures~\ref{Fig:Scatter}~and~\ref{Fig:HaloMass}, except
the infall rates are plotted against  the stellar mass of each parent halo. 
 As can be seen from the \textit{left panel}, the \rarepeak region has a large fraction of
  metal-free haloes residing in low stellar-mass haloes. A substantial
  fraction of these haloes have high infall rates. The extremely metal poor haloes tend to have
  larger larger stellar masses and the metal-enriched haloes more stellar mass still as expected. 
  The \textit{right panel} shows that the \normal region has significantly fewer metal-free haloes
  but that they nonetheless populate the high mass infall region of the plot. }
 \label{Fig:StellarMass}
\end{center} \end{minipage}

\end{figure*}
\begin{figure*}
\centering
\begin{minipage}{175mm}      \begin{center} 
\centerline{
\includegraphics[width=0.35\textwidth]{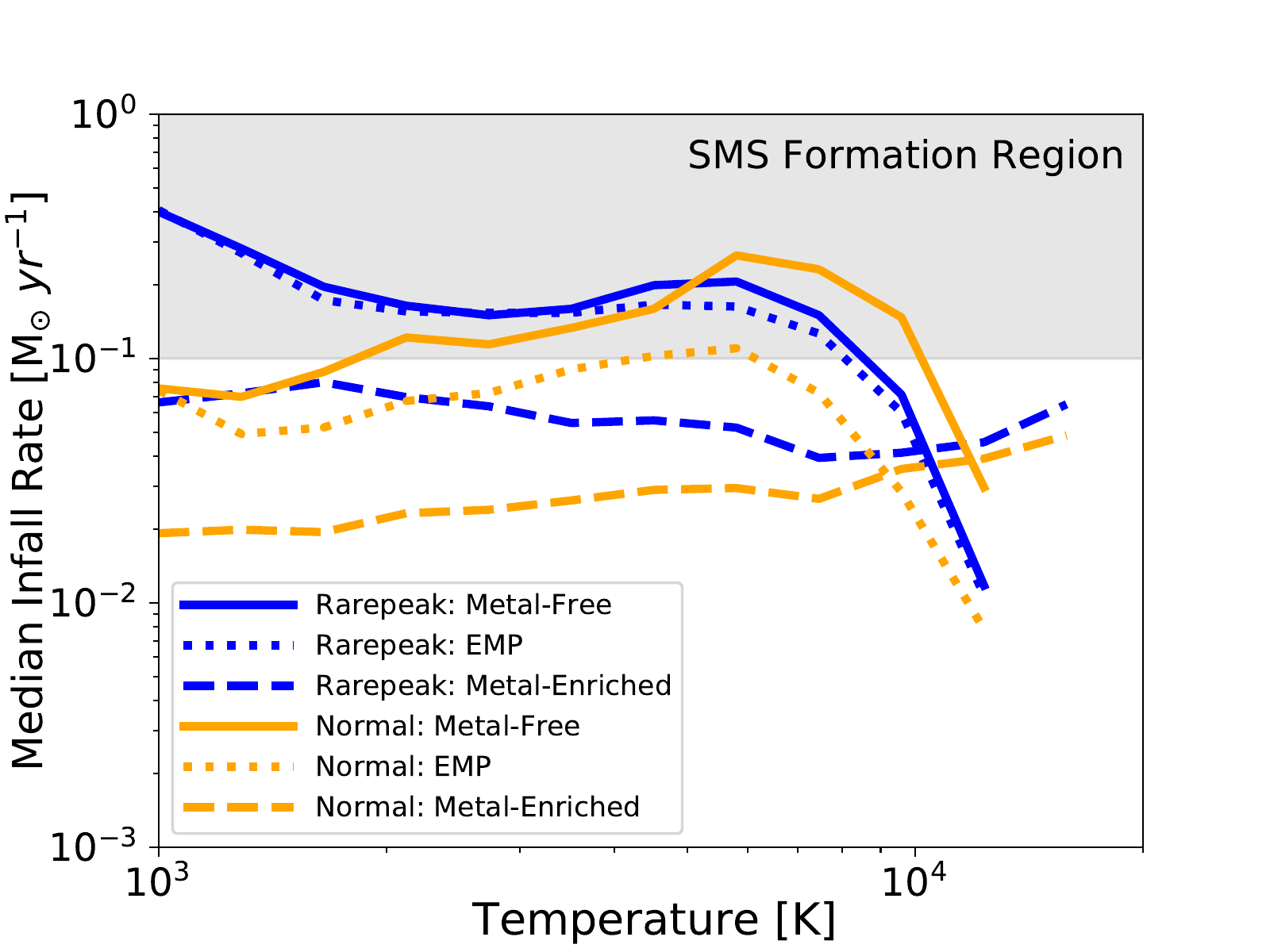}
\includegraphics[width=0.35\textwidth]{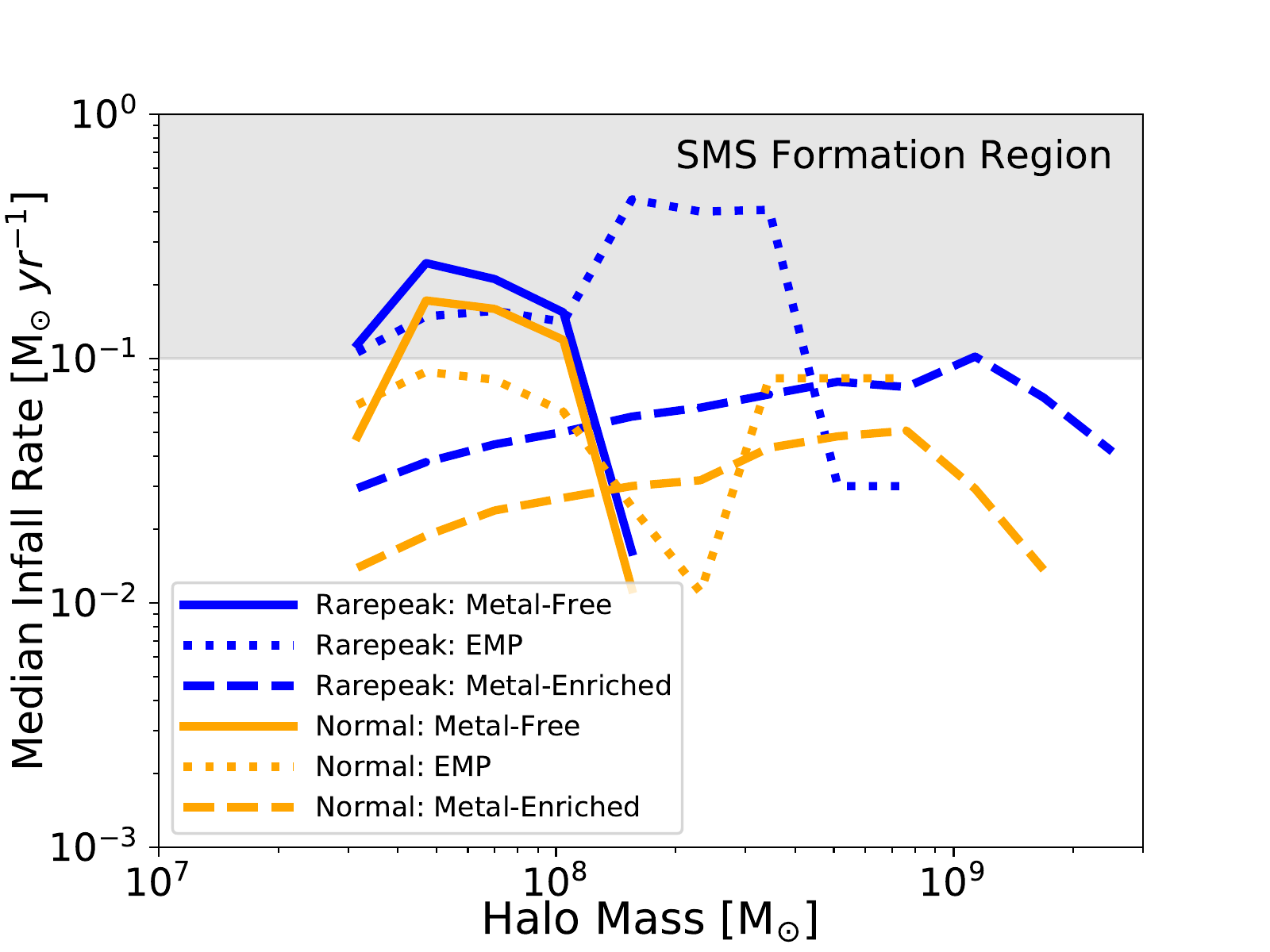}
\includegraphics[width=0.35\textwidth]{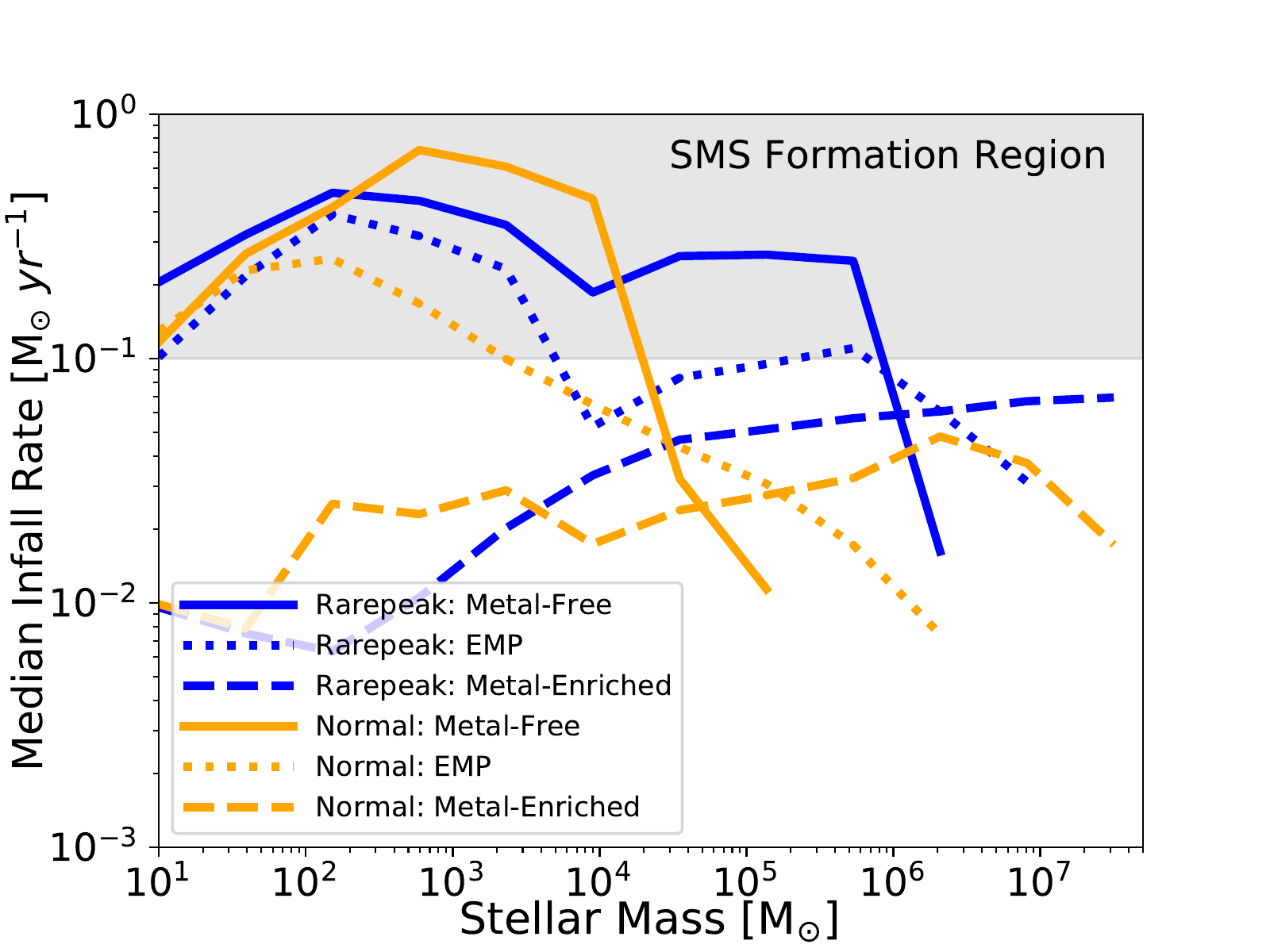}}
\caption{In the three panels, we show how the median infall rate within 20 pc of the centre of each halo
  varies with temperature (left panel), halo mass (middle panel) and stellar mass (right panel). The metal-free
  haloes are shown by solid, the EMP haloes by dotted, and the metal-enriched haloes by dashed lines.
  In all cases the metal-free and EMP haloes show the highest median infall rates for the regions where they
  exist. The \rarepeak region also has systematically higher infall rates compared to the \normal region.
  In each panel, the median rates vary between approximately 0.01 \msolaryr and 1.0 \msolaryrc. High
  central infall rates are found in metal-free and EMP haloes clustered around the atomic cooling limit (i.e.
  a few times $10^7$ \msolarc) but towards lower stellar masses. This breakdown of central infall rate
  by metallicity shows that up to $Z \sim 10^{-3}$ \zsolar central infall rates are likely to be sufficient
  to significantly modify the IMF of stars in a halo with the possibility of SMS formation. The grey shaded
  region above 0.1 \msolaryr signifies the region in which SMS formation may be possible.
 \label{Fig:BinnedMedianInfall}}
\end{center} \end{minipage}

\end{figure*}
\subsection{Central mass infall rates}
\begin{figure*}
\centering
\begin{minipage}{175mm}      \begin{center} 
\centerline{
\includegraphics[width=0.525\textwidth]{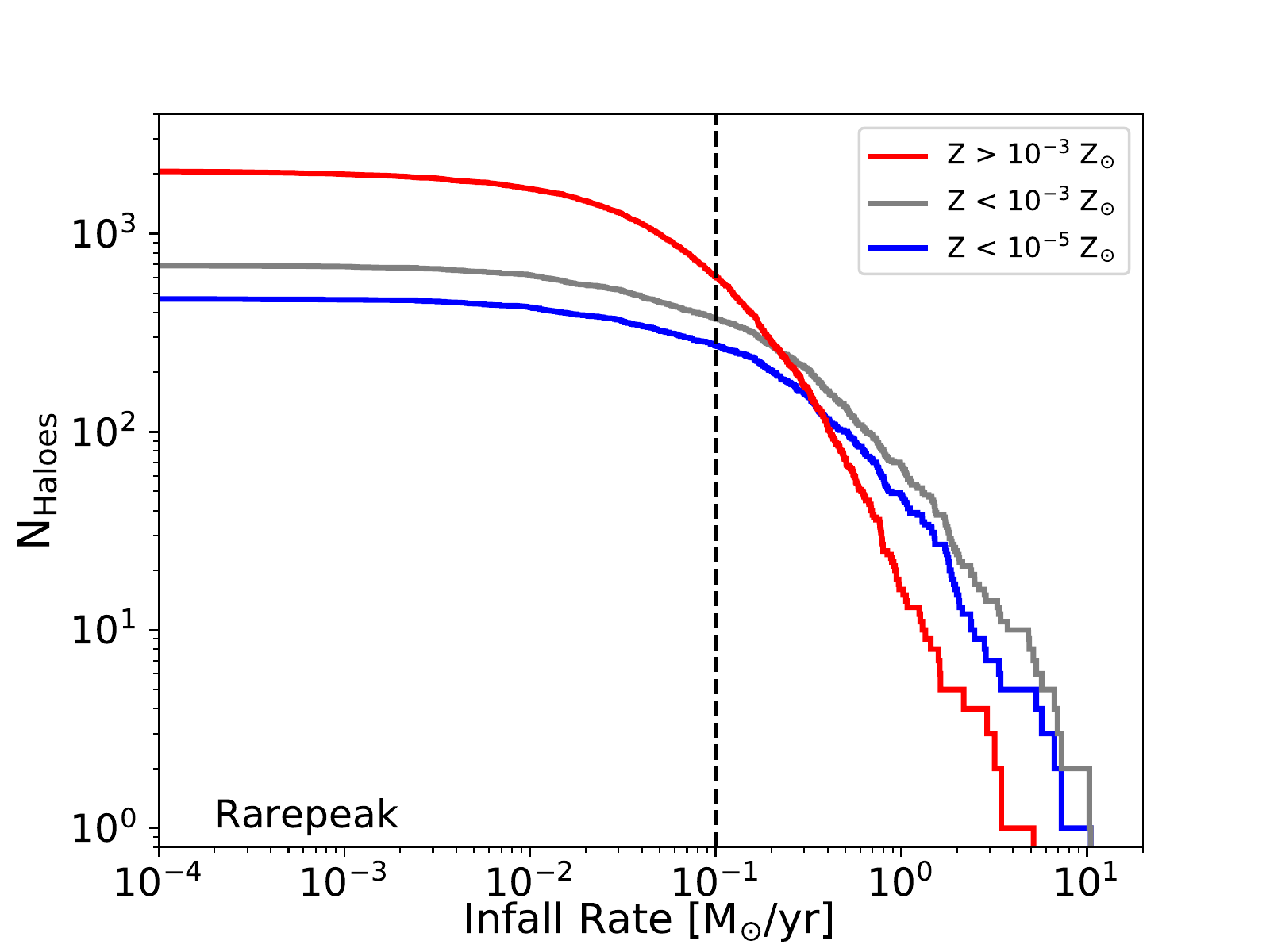}
\includegraphics[width=0.525\textwidth]{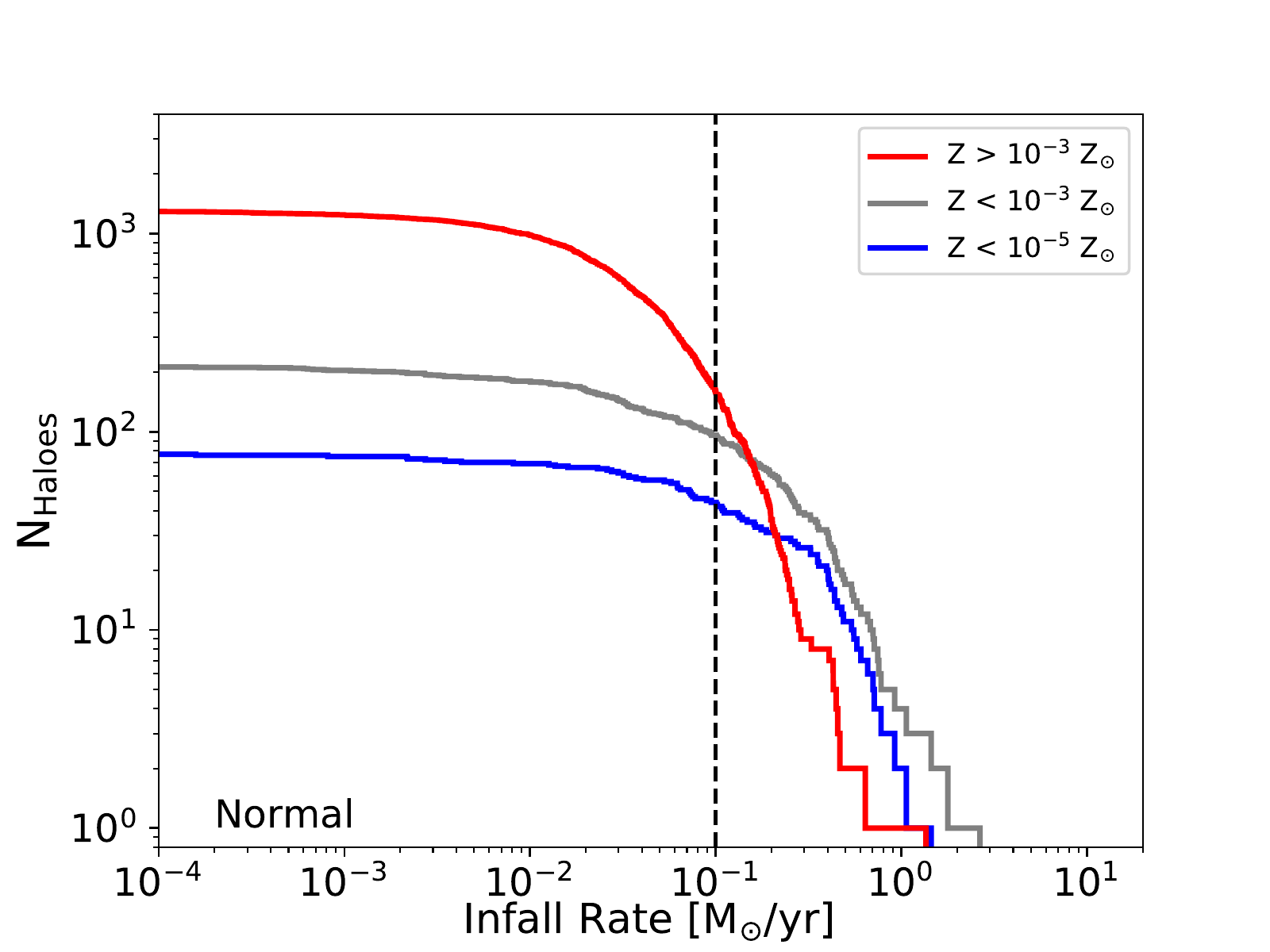}}
\caption{\textit{Left Panel}: The cumulative number of haloes with infall rates greater than the given
  x-axis value for the \rarepeak region.  The total number of \rarepeak haloes across all outputs
  is 3,244. Infall rates are again calculated at 20 pc from the halo centre.
  Haloes are grouped into haloes with metallicities $Z> 10^{-3}$ \zsolar (red line),
  $Z< 10^{-3}$ \zsolar (grey line) and  $Z< 10^{-5}$ \zsolar (blue line). At mass infall
  rates greater than 0.1 \msolaryrc, the number of metal-enriched haloes is approximately 2.2 times
  higher than the number of 
  metal-free haloes ($Z< 10^{-5}$ \zsolarc). Therefore, if metal-enriched haloes can support SMS
  formation then the number density of candidate haloes can increase substantially. 
  \textit{Right Panel}: The distribution of
  infall rates for the \normal region. The  total number of \normal haloes across all outputs is 2,334.
  In the \normal region at mass infall
  rates greater than 0.1 \msolaryr there is up to a factor of four more metal-enriched haloes compared to
  metal-free haloes and up to twice as many EMP haloes compared to metal-free haloes. 
 } \label{Fig:Histogram}
\end{center} \end{minipage}

\end{figure*}
\indent We first plot the mass infall rate of each halo against three different halo
characteristics. First in Figure~\ref{Fig:Scatter} we show a scatter plot of the mass infall rates
against the inner halo temperature. Secondly in Figure~\ref{Fig:HaloMass} we plot the halo infall
rate against the total halo mass, and finally in Figure~\ref{Fig:StellarMass} we show the halo
infall rate against the stellar mass of each halo. We begin by analysing Figure \ref{Fig:Scatter}. 
In the left panel of Figure \ref{Fig:Scatter} we show the results from the \rarepeak
simulation and in the right panel from the \normal simulation. Each point shows the mass
infall rate averaged over the inner 20 pc of each atomic cooling halo for the \rarepeak
and \normal region. The mass infall rates are plotted against the average\footnote{All
  averages are computed using a cell volume weighted mean} temperature within
the same region in each halo. There are two colour codes on the plot. The grayscale
hexbins capture all metallicities\footnote{Note that all metallicities
  referred to in this paper are gas phase metallicities.}
greater than $Z = 10^{-3}$~\zsolarc. Each coloured circle then refers to
the metallicities of the EMP and metal-free haloes. Red circles indicate EMP haloes
 ($10^{-5}$~\zsolar $ \lesssim Z \lesssim 10^{-3}$~\zsolarc)
while blue circles indicate metal-free regions  ($Z \lesssim 10^{-5}$ \zsolarc). We also
include green circles on the plot to indicate haloes which are metal-free and star-free. 

Some results are immediately clear from a comparison of the panels in Figure~\ref{Fig:Scatter}.
The \rarepeak region clearly has significantly more haloes that have lower
metallicities (and/or are metal-free) and it also has more haloes
with high infall rates. While this is true by number it is also true by fraction.
For the \rarepeak region $\sim 25$\% are classified as EMP or metal-free compared to
less than 15\% for the \normal region. The \rarepeak region is also characterised by a
large number of rapidly collapsing, EMP and metal-free haloes. 
The metal-free and star-free haloes (green circles) show a strong scatter and are not preferentially
rapid accretors. However, the metal-free and star-free haloes do populate the lower-temperature parts of the
phase space. This is because these haloes have no ongoing star formation (by definition) and so
the temperature of the gas is determined purely by the competition between gas cooling and
dynamical and compressional heating. The other EMP and metal-free haloes, especially those with gas temperatures
close to T = $10^4$ K are being heated by ongoing star formation with atomic line
cooling keeping the temperature close to, or slightly below, T = $10^4$ K. 
The \normal region, as mentioned above, is dominated by metal-enriched haloes with over 85\% of haloes being metal-enriched.
Furthermore, the fraction of rapidly collapsing haloes is greatly diminished compared to the \rarepeak region. In the
\normal region 10\% are rapidly collapsing (dM/dt $>$ 0.1 \msolaryrc) compared to 30\% for the \rarepeak region.
In Figure~\ref{Fig:HaloMass} we plot the mass infall rate against the total (dark matter
and gas) halo mass. In this representation the EMP and metal-free haloes are strongly clustered, in both
the \rarepeak and \normal regions, at the atomic cooling limit (M$_{\rm atm} \sim 2 - 5 \times 10^7$
\msolarc). This result is not surprising; once the atomic cooling limit is reached, star formation
will begin (if it hasn't already), and from that point onwards, it is only a matter of time before
the halo becomes metal enriched. There is also a noticeable decline in the number of high infall-rate
haloes towards high halo masses. This trend is also seen in Figure \ref{Fig:BinnedMedianInfall} (see below).
This can be explained through the natural halo merging process where due to the exponential form of the halo mass
function smaller haloes are more populous than larger haloes. What we see in Figure \ref{Fig:HaloMass} is that mergers of smaller haloes
into more massive haloes is driving high mass inflow rates for halos below approximately $10^{8}$ \msolarc. As we increase the mass
scale the number (distribution) of haloes becomes smaller (larger) and hence the mass inflow rates decline. This point adds to
the fact that haloes just transitioning into the atomic cooling regime are ideal sites for SMS formation. \\
\indent Complementary to this is Figure~\ref{Fig:StellarMass}, 
where we plot the mass infall rate against the stellar mass. In the left panel of Figure~\ref{Fig:StellarMass}
we show the scatter/hexbin plot, coloured by metallicity, of the
\rarepeak region. There is a large cluster of metal-free haloes, with $Z \lesssim 10^{-5}$ \zsolarc,
with stellar masses of M$_{*} \lesssim 10^3$ \msolarc. These haloes formed PopIII
according to the sub-grid PopIII star formation prescription with masses dictated 
by a pre-defined initial mass function. However, given the rapid infall rates these 
haloes could also have formed SMSs had such a sub-grid prescription existed. The
haloes are identified here however as their supernovae have not yet fully enriched the 
halo or the stars have not exploded yet. Allied to this 
it should also be noted that this group of haloes also contain some of the highest mass infall
rates. With increasing stellar mass the halo metal enrichment levels increase and most EMP haloes
are found with stellar masses scattered between $10^{3}$ \msolar and $10^{6}$ \msolarc. The metal-free haloes
do however have a bias towards higher mass infall rates compared to the EMP haloes.
The \normal region contains significantly fewer metal-free haloes and
similar to the \rarepeak region, the majority have stellar masses
with M$_{*} \lesssim 10^3$ \msolarc. The overall trend however, while less pronounced than the \rarepeak region, is
similar. \\
\begin{figure*} [!t]
\centering
\begin{minipage}{175mm}      \begin{center} 
\centerline{
\includegraphics[width=0.525\textwidth]{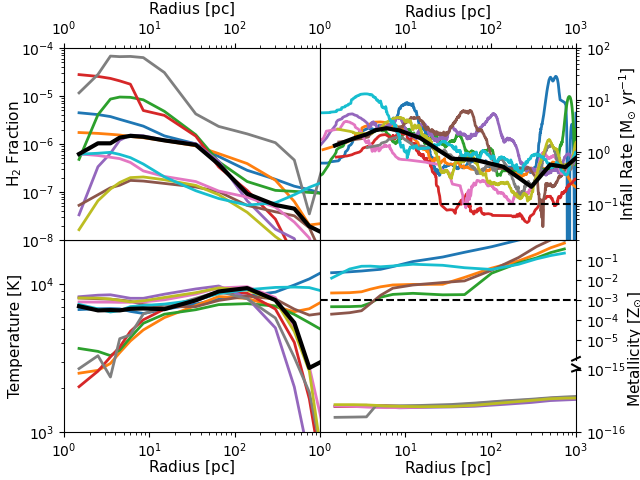}
\includegraphics[width=0.525\textwidth]{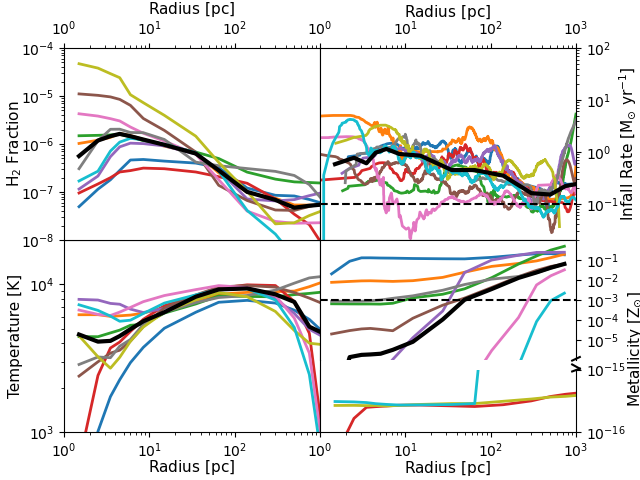}}
\caption{\textit{Left Panel}: Radial profiles for the top 10 highest infall-rate haloes
  in the \rarepeak region. For
  each halo the temperature, \molH fraction, infall rate and metallicity are given for
  each halo. The median radial values for each quantity are shown using the
  thick black line in each panel. The metallicity (bottom right panel) of haloes can span many
  orders of magnitude and so that panel is artificially cut in the middle. Nonetheless, what is
  clear is that the highest infall haloes are evenly spread between those which are metal-free and
  those which are metal enriched. However, note that even for the metal-enriched haloes the
  metallicity decreases in all cases towards the halo centre. 
  \textit{Right Panel}: The same plot for the \normal
  haloes. The features are broadly similar, the metallicity sub-panel shows more haloes in transition
  compared to the \rarepeak haloes with a number of haloes showing strongly declining metallicities.
  Note also that in all cases the haloes have mass infall rates which exceed the critical value
  for SMS formation of 0.001 \msolaryr \citep{Haemmerle_2018} while also exceeding
  0.1 \msolaryr at almost all radii (dashed black line).
  Note that values of the metallicity (bottom right sub-panel) which fall below $Z \sim 10^{-15}$ \zsolar should be
    treated as zero.
} \label{Fig:RadialProfiles}
\end{center} \end{minipage}

\end{figure*}
\indent In Figure \ref{Fig:BinnedMedianInfall} we show the median infall rate for all haloes as
  a function of temperature, halo mass and stellar mass respectively. Within each panel, the central
  infall rates are further broken down according to metallicity. For each case the median
infall rate is calculated as the median infall rate within 20 pc of the halo centre. To plot the
median of all haloes these median values are then binned against temperature, halo mass and
stellar mass and the median of each bin again calculated. Haloes are further subdivided by their
metallicity as shown. In each case we see that the \rarepeak region has a near
systematically higher infall rate reflecting the more clustered environment of the \rarepeak
region with mergers and halo interactions more common \citep{Wise_2019, Regan_2020}. \\
\indent However, what
is also very noticeable is that the metal-free and EMP haloes have significantly higher central
infall rates compared to the metal-enriched haloes. The high infall rates suppress early star
formation with star formation only taking hold once the halo mass exceeds the atomic cooling limit
(see middle panel). The high infall rate haloes cool in the centre due to
\molH line cooling so their temperatures extend down to 1000 K or below. Many of these
(high infall rate) haloes have small stellar masses peaking around $10^3$ \msolarc (see also
Figure \ref{Fig:StellarMass}). A large number of these haloes, with high infall rates close to the
atomic cooling limit, are strong candidates for SMS formation. It should also be noted that at
higher halo masses (and stellar masses) that the different metallicity lines essentially converge.
The higher halo masses have generally higher metallicities and stellar content. Within these haloes,
if SMS formation is to be realised, inhomogeneous pockets of lower metallicity gas will be required.
\\
\indent In summary, from Figures~\ref{Fig:Scatter}, \ref{Fig:HaloMass}
and \ref{Fig:StellarMass} we see that there exists a sizeable minority of rapidly collapsing haloes
which are metal-free ($Z < 10^{-5}$ \zsolarc). These
haloes, due to their large mass infall rates, makes them strong candidates for supermassive
star formation \citep{Woods_2018, Chon_2020}. However,  there exists an even larger population of
rapidly collapsing, extremely metal-poor and metal enriched haloes. Some of these haloes could
potentially support SMS formation \citep[see e.g.][]{Chon_2020},
provided the infall rate does not significantly decrease further inward; 
and for the metal-enriched haloes if the metal mixing within the halo is inhomogeneous. We will discuss this point further in \S
\ref{Sec:Discussion}. First we quantify the relative abundance of metal-free, extremely metal-poor
and metal-enriched haloes
in the next section.

\subsection{The abundance of metal-free, extremely metal-poor and metal-enriched haloes}
In Figure~\ref{Fig:Histogram} we show the cumulative number of haloes with infall rates greater than
the value given on the x-axis. The x-axis runs from
$10^{-4}$ \msolaryr $\leq \dot{M}(R=20~{\rm pc}) \leq 10^{1}$ \msolaryrc. 
 The mass infall rate,
$\dot{M}(R)$, is calculated as the volume averaged value within 20 pc of the centre for both the \rarepeak 
and \normal regions. The
\rarepeak region is shown in the left panel while the \normal region is shown in the right 
panel. Lines in each panel are coloured as follows: red lines are for metal-enriched haloes with
metallicities ($Z > 10^{-3}$ \zsolarc), grey lines are for haloes with ($Z < 10^{-3}$ \zsolarc) and
blue lines are for haloes with ($Z < 10^{-5}$ \zsolarc).\\
\indent Concentrating first on the \rarepeak region, we see that there are a small number of
haloes ($84$, or 2\% of the total) with mass infall rates greater than 1 \msolaryrc. Among these (very high-accreting)
haloes most are either EMP or metal-free ($\sim 81$\%). However, what is also very relevant from this figure
is the number of haloes that occupy the EMP range (i.e. difference in height between the
blue and grey lines). If we focus on all haloes with mass infall rates
exceeding\footnote{As previously noted the mass accretion rate onto a protostar must exceed
  0.001 \msolaryr \cite{Haemmerle_2018}, however, we choose a threshold here of 0.1 \msolaryr
  because we are measuring not the
  stellar accretion rate but the central mass infall rate at 20 pc.}  0.1
\msolaryr
then accounting for EMP
haloes this increases the number of SMS candidates by a factor of 1.4 and allowing all metal-enriched haloes
with mass infall rates greater than 0.1 \msolaryr increases the number of SMS candidate haloes
by more than a factor of two
compared to the metal-free case. We will return to this point in \S \ref{Sec:Discussion} below. \\
\indent In the right panel of Figure~\ref{Fig:Histogram},
we show the same histogram for the \normal region. In this case there are two metal-free haloes
and two EMP haloes with mass infall rates greater than 1 \msolaryr. There is only a single metal-enriched
(with a metallicity value of 0.0015 \zsolarc)
halo with the same mass infall rate. In the \normal region the haloes are dominated by metal-enriched haloes
for mass infall rates greater than 0.1 \msolaryr
On the face of it this suggests that the \normal region, similar to what was found in the \rarepeak region, will
support only a small number of supermassive star candidates. However, again
if we relax the criteria for supermassive star formation and consider also EMP and metal-enriched haloes
then the number of possible candidate haloes can increase by more than a factor of two when we include EMP haloes
and up to a factor of four when we include metal-enriched haloes. \\
\indent In summary what we find is that by including EMP haloes, which \cite{Chon_2020} and
\cite{Tagawa_2020} have recently found do support SMS formation, the number of SMS formation haloes
can be increased by a factor of at least two. Extending this further to include haloes
with metallicities greater than $Z \sim 10^{-3}$ \zsolar could potentially increase the number of
SMS formation sites by a factor of up to four or more compared to the metal-free case. Alternatively,
such rapidly collapsing haloes may provide the ideal environment to support the growth of light seed
black holes \citep[e.g.][]{Alexander_2014, Inayoshi_2018}.
\begin{figure*}
\centering
\begin{minipage}{175mm}      \begin{center} 
\centerline{
\includegraphics[width=0.525\textwidth]{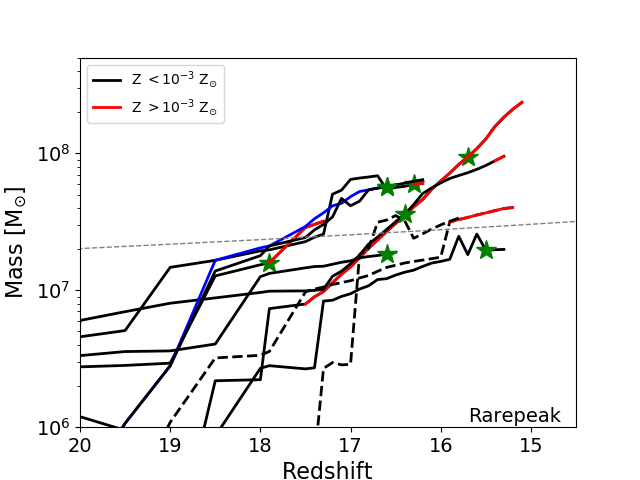}
\includegraphics[width=0.525\textwidth]{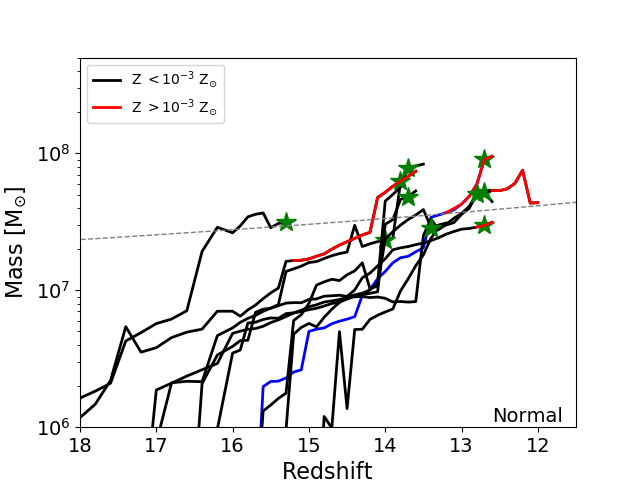}}
\caption{\textit{Left Panel}: The growth rate as a function of redshift for the \rarepeak haloes
  identified in the left
  panel of Figure \ref{Fig:RadialProfiles}. Metal-free and extremely metal-poor ($Z < 10^{-3}$
  \zsolarc) haloes are marked in black, while metal-enriched haloes ($Z > 10^{-3}$ \zsolarc) are
  marked in red. Note that a halo can exit the metal-free regime during its evolution. The
  enrichment can come from internal or external sources. Green stars indicate the redshift at
  which star formation first occurs in each halo. Two haloes have remained star-free until their
  detection as high infall-rate haloes (indicated by dashed curves). The blue line is the halo which displays the
  highest instantaneous mass infall rate. 
  The dashed gray near-horizontal line in each panel is the atomic cooling limit.
  \textit{Right Panel}: The same mass growth plot for the \normal region. The trend is similar
  here with an almost even split between haloes which remain extremely metal-poor or metal-free and
  those which become enriched. This demonstrates that metal-enriched haloes can have
  mass infall rates comparable to those which are metal-free all the way into their centre. 
}\label{Fig:GrowthRates}
\end{center} \end{minipage}

\end{figure*}
\subsection{The Most Rapid Accretors}
\indent We next examine the ten most rapidly accreting haloes, when ranked independent of their metallicity, in both simulated regions.
In Figure~\ref{Fig:RadialProfiles} we plot the radial profiles of each of these rapidly
accreting haloes.
In each panel, we show the temperature, \molH fraction, mass infall rate and
metallicity of each halo as a function of radius. 
The thick black line in each panel gives the
median value for the given quantity. Values across the two panels are broadly similar. Both regions
show gas-phase metallicity values that range from completely metal-free (Z $\sim 10^{-16}$ \zsolarc)
into the metal-free regime and all the way through to those which are metal-enriched (Z $> 10^{-3}$
\zsolarc). Note that the metallicity sub-panel of each panel
is split between approximately extremely metal-poor and completely metal-free haloes. The dashed
line in the metallicity panel indicated the critical metallicity above which SMS formation
has been found by \cite{Chon_2020} to be precluded to occur. Below this line SMS is possible if the
accretion rate is sufficient. Note also that each radial profile of each halo's  metallicity shows
decreasing metallicity towards the centre of the halo. For some haloes this decrease is highly
pronounced, particularly for the \normal region. These profiles give some evidence that metal
mixing within haloes is likely to be somewhat inhomogeneous and incomplete.\\
\indent The mass infall rate is shown in the top right panel of each figure. 
Mass infall rates in both the \rarepeak and \normal regions, for these select haloes, are
comfortably above the critical value thought to be necessary for SMS formation (0.001
\msolaryrc; \citealt{Haemmerle_2018}). The dashed black line marks the value of 0.1 \msolaryr which we use to demarcate very high infall rate haloes.
While mass infall rates determined at this radius are in no
way a guarantee that this will lead to similar accretion rates onto a protostar, the high rates do
suggest that large amounts of matter are being transferred to the halo centre at least. \\
\indent In Figure~\ref{Fig:GrowthRates} we plot the halo mass history of the highest
infall-rate haloes shown in Figure~\ref{Fig:RadialProfiles} with the \rarepeak and \normal haloes
shown on the left- and right-hand panels, respectively. In the panels we identify haloes which
are metal-enriched (Z $> 10^{-3}$ \zsolarc) and those below that threshold. The growth of
metal-enriched haloes are marked in red while those below the threshold are marked in black. Note
that haloes often transition from EMP/metal-free to metal-enriched during the course of their growth. 
Haloes can become metal-enriched either through internal star formation and
supernovae or alternatively can become metal enriched through external enrichment processes
\citep[e.g.,][]{Smith_2015}. Where star formation first
occurs in a halo, the redshift of star formation is denoted by a green star on the halo growth line. 
Haloes which remain completely star-free are marked as dashed lines. Finally, the blue line denotes
the halo which shows the highest instantaneous infall rate in the centre. \\
\indent In the \rarepeak region there are two
star-free haloes. One of these haloes becomes metal-enriched, although it has yet to form
a star by the time the halo is detected as a high infall-rate halo. The time of metal enrichment
compared to star formation is somewhat varied between haloes. Some haloes form stars and then
metal-enrichment follows (internal enrichment). In other cases, metal-enrichment precedes
star formation, as noted above.
However, given that these galaxies have deeper potential wells and longer sound crossing times,
the metal mixing can be gradual in some haloes but more rapid in others as was shown in Figure~\ref{Fig:RadialProfiles}.
The \normal region haloes in the right hand panel show similar trends.
The \normal region has no star-free haloes, with all haloes in this subset undergoing at
least some star formation.

\section{Discussion and Conclusions} \label{Sec:Discussion}

The goal of this paper is to examine the prevalence of high infall-rate halos in the
Renaissance Simulations, which in turn are thought to accurately represent early structure formation in our Universe \citep{Chen_2014, Xu_2013, Xu_2014, OShea_2015,
  Barrow_2017, Wise_2019}. A high mass accretion rate onto an embryonic protostar is thought to be the
single most important criterion for the formation of a SMS \citep{Hosokawa_2013, Sakurai_2016,
  Woods_2018, Haemmerle_2018}. Therefore, we examine the Renaissance datasets for haloes which are experiencing
rapid gas inflow in their centres. While rapid gas inflow to the halo centre, 
measured at 20 pc from the centre, in order to ensure good cell sampling,
does not guarantee
that a SMS will ultimately form, it is very likely to be a necessary condition. \\
\indent Previous investigations of SMS candidate haloes\footnote{These haloes are often
  referred to as "direct collapse black hole" host haloes in the literature due to the fact
  that the high infall rates are thought to lead to massive black hole formation in certain
  circumstances. However, current research indicates that the intermediary stage involves a
  SMS, with no supernova event, and hence we use that terminology here.} have almost
universally focused on
metal-free haloes in which the equation of state of the gas was determined by the gas phase
cooling mechanisms of hydrogen and helium. Here, we extend the analysis to haloes that are either
extremely metal-poor ($10^{-3}$ \zsolar $< Z < 10^{-5}$ \zsolarc) or metal-enriched ($Z > 10^{-3}$ \zsolarc). If the number density of haloes that support SMS formation
can be extended to include non metal-free haloes (at least perhaps up to some
metallicity or due to metal enrichment inhomogeneity) then the
number density of candidate SMS host haloes will increase accordingly, and so is of
immense interest to the community. 
Previous SMS candidate halo number densities, including metal-free haloes only,
have tended to show that the number of SMS candidate haloes may be just sufficient to explain the number density of high-z quasars,
not leaving much room to lose a large fraction of these candidates~\citep{Agarwal_2012, Visbal_2014b, Agarwal_2015b, Latif_2014a,
  Valiante_2016, Habouzit_2016, Valiante_2017, Habouzit_2017, Regan_2017}. \\
\indent \cite{Chon_2020} have recently shown, through dedicated high-resolution numerical simulations
of the collapse of candidate SMS haloes that metallicities with $Z < 10^{-3}$ \zsolar
are compatible with SMS formation. They showed that for metallicities of $Z \lesssim 5 \times 10^{-5}$ \zsolar
the behaviour of SMS formation is the same as the metal-free case, i.e. SMS formation results
directly from the rapid accretion of pristine gas. 
\cite{Chon_2020} further showed that when the metallicity enters the
extremely metal-poor regime, $5 \times 10^{-5}$ \zsolar $ < Z < 10^{-3}$ \zsolarc, then while
the gas does fragment, the fragments nonetheless coalesce onto the growing protostar and the
result is a SMS. 
\cite{Tagawa_2020} came to similar conclusions
using a semi-analytic approach (see also \citealt{Inayoshi_2014b} for similar conclusions in the case of a fragmenting disc in an atomic-cooling halo).
In the case of higher metal enrichment,  $Z \gtrsim 10^{-3}$ \zsolar,
\citet{Chon_2020} find that gas cooling becomes too rapid and the gas fragments into low-mass fragments
and SMS formation is inhibited. However, there remains the possibility even in these
metal-enriched haloes that pockets of metal-poor gas will remain in which SMS formation can occur.
The topic of metal inhomogeneity was outside the scope of  \cite{Chon_2020} 
but given that metal mixing has previously been shown to have very strong mixing gradients
in haloes in the early universe
\citep[e.g.][]{Smith_2015} this (metal-enriched) scenario warrants further research, similar to 
what \cite{Tarumi_2020} have done in the context of stellar archaeology where they looked at the
difference in metal mixing between internally and externally enriched galaxies. \\
\indent With this in mind, we included all rapidly collapsing haloes in our analysis, regardless of
metallicity. To do this, we examined the Renaissance datasets, analysing
both the \normal and \rarepeak regions. We analysed the simulation volumes and filtered atomic
cooling haloes by their instantaneous mass infall rates, averaged within 20 pc of the centre of each halo.  \\
\begin{figure}
   \centering 
\includegraphics[width=0.525\textwidth]{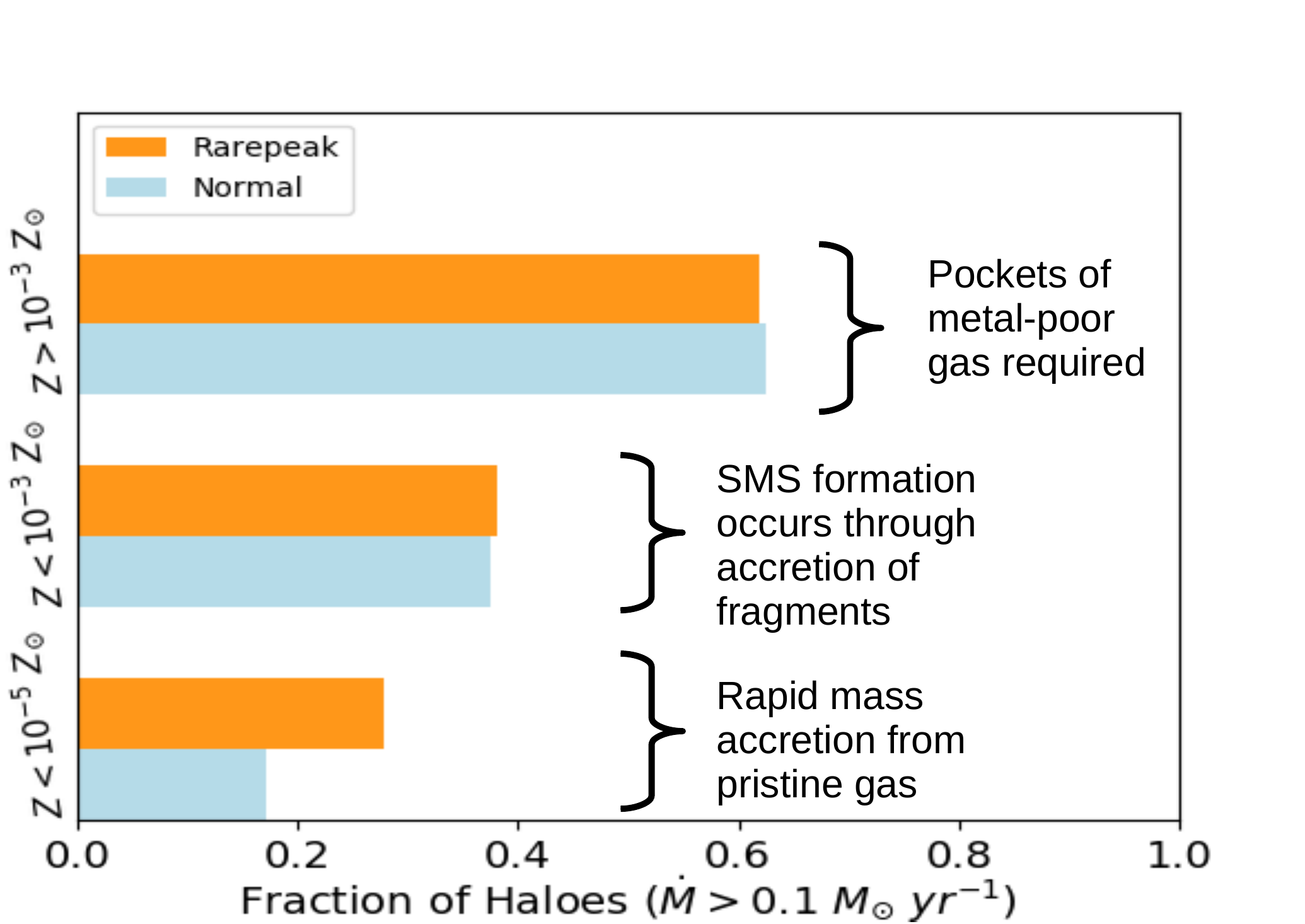}
\caption{The fraction of haloes which have metallicities of $Z > 10^{-3}$~\zsolar,
  $Z < 10^{-3}$ \zsolar and $ Z < 10^{-5}$ \zsolar and have mass infall rates greater than
  0.1 \msolaryrc. \rarepeak haloes are coloured in orange. \normal haloes in blue. 
  The majority of haloes in both cases are metal enriched and hence within
  those haloes pockets of extremely metal-poor gas 
  is likely required, including on scales below those resolved here, for SMSs to form. } \label{Fig:Fractions}
\end{figure}
\indent We found that, as expected, there is a large number of haloes with
very high central infall rates (see Fig.~\ref{Fig:Histogram}). At infall rates
greater than 1 \msolaryr the fraction of rapidly collapsing haloes are dominated by haloes with
metallicities less than $10^{-3}$ \zsolar ($\sim 81$\%).
Relaxing the
mass infall criteria to 0.1 \msolaryrc,
we find that among these ``high-accreting''
haloes, the majority are either metal-enriched ($Z > 10^{-3}$ \zsolarc) or
extremely metal-poor ($10^{-5}$ \zsolar $ < Z < 10^{-3}$ \zsolarc). \\
\indent In Figure \ref{Fig:Fractions}
we show the fraction of high accretion rate ($\dot{M} > 0.1$ \msolaryrc) haloes and
their (gas phase) metallicities for haloes from both the \normal and \rarepeak simulations.
The \rarepeak haloes are displayed in orange, the \normal haloes in blue. For the \rarepeak
region the majority (62\%) are metal-enriched (i.e. $Z > 10^{-3}$ \zsolarc) with 38\% extremely
metal-poor or metal-free. For the \normal region the fractions are very similar.
Since the \normal region represents a typical
region of the Universe, the results for the \normal region can be assumed representative of 
regions likely to be observed, at least by upcoming wide-field surveys (although not quite
reaching yet the high redshifts probed in this study, e.g. Nancy Grace Roman Space Telescope and
Euclid). While a comprehensive study of the number density of SMS candidate haloes is outside the
scope of the present paper, we estimate the number density of SMS candidate
haloes as a function of metallicity by examining the final snapshots for the \rarepeak and \normal
simulations. If we again consider haloes with central mass infall rates greater than $0.1$
\msolaryrc, then we find that for the \rarepeak region the number density of metal-free haloes
is 0.26 cMpc$^{-3}$. Including the EMP haloes increases the number density to 0.35 cMpc$^{-3}$.
Recall that the \rarepeak environment is a biased region and so the number densities found here
should be multiplied by a factor between $10^{-3} - 10^{-4}$ to account for this bias \citep{Wise_2019}.
Doing this we find a number density of approximately $10^{-5}$ cMpc$^{-3}$ as a global estimate.
This  number density is nonetheless orders of magnitude above the observed number density of
SMBHs at z = 6 (number density $\sim 1$ cGpc$^{-3}$). We are currently investigating the wider number
density implication of our calculations (Arridge et al. in prep) but this preliminary result
confirms and reinforces the result found in \cite{Wise_2019} that there is likely to be a
large population of faint quasars/intermediate mass
black holes at high redshifts.
\\
\indent In the optimistic case, where all of these metal-enriched haloes support SMS formation, the
boost to
candidate haloes for SMS formation, compared to the metal-free case, can be increased by a factor of
at least four. Some of these haloes may ultimately be shown through both  higher-resolution simulations
and through considering additional physics (e.g. dust and gas opacity) 
to not support SMS formation. Nevertheless, their high mass infall rates will, at a minimum, transfer
a large amount of baryons towards the halo centre, which may impact the initial mass function and
is worth further investigation. Furthermore, such high-infall rate haloes may support the rapid
growth of less massive, pre-existing, black holes. \\
\indent Additional high-resolution hydrodynamic simulations will be required to
determine the fraction of metal-enriched haloes which can support SMS formation, and, in particular,
whether the accretions rates in the cores remain sufficiently high.  This could occur if pockets
of metal-poor gas in the cores of these haloes remain, allowing rapid accretion and supporting SMS
formation. There are clear indications from our analysis of the radial metallicity profiles
(see Figure \ref{Fig:RadialProfiles}) that metal mixing decreases strongly towards the halo centre.
We ran an additional analysis of the SMS candidate haloes using a narrower prescription for the
radius at which we calculate the accretion rates and the metallicity. Using a radius of 10 pc
from the centre the number of EMP and metal-free haloes detected increases to 58\% up from 38\%
reflecting the strong metallicity gradient found in most haloes. 
If this mixing inefficiency persists to even smaller radii (which it appears to),
then metal inhomogeneity may be common in the cores of these young proto-galaxies. The galaxies
studied here are embryonic and have had very few dynamical times in which 
gas mixing can take place and hence extremely metal-poor pockets are much more likely here than
in more mature and evolved galaxies at lower redshift.\\
\indent  The addition of extremely metal-poor and metal-enriched haloes to the number density of SMS
candidate haloes would help relieve the current tensions that exist in understanding the formation
pathways for high-$z$ quasars. More generally, they could significantly modify the
``initial mass function" of the earliest supermassive black holes.

\section*{Acknowledgments}

\noindent JR acknowledges support from the Royal Society and Science Foundation Ireland under
grant number URF$\backslash$R1$\backslash$191132.
ZH acknowledges support from NASA grant NNX15AB19G and National Science Foundation grant AST-1715661.
JHW is supported by National Science Foundation grants AST-1614333 and
OAC-1835213, and NASA grants NNX17AG23G and 80NSSC20K0520.  
BWO acknowledges support from NSF  grants  PHY-1430152,  AST-1514700, AST-1517908, and
OAC-1835213,  by  NASA grants NNX12AC98G and NNX15AP39G, and by HST-AR-13261 and HST-AR-14315.  
The Renaissanace simulations were performed on Blue 
Waters, which is operated by the National Center for Supercomputing Applications (NCSA)
with PRAC allocation support by the NSF (awards ACI-1238993, ACI-1514580, and OAC-1810584).
This research is part of the Blue Waters sustained-petascale computing project, which
is supported by the NSF (awards OCI-0725070, ACI-1238993) and the state of
Illinois. Blue Waters is a joint effort of the University of Illinois at
Urbana-Champaign and its NCSA.  The freely available plotting library {\sc
matplotlib} \citep{matplotlib} was used to construct numerous plots within this
paper. Computations and analysis described in this work were performed using the
publicly-available \enzo \citep{Enzo_2014, Enzo_2019} and \yt{} \citep{YT} codes, which are the product of a
collaborative effort of many independent scientists from numerous institutions
around the world. Their commitment to open science
has helped make this work possible. Finally, the authors would like to
thank the anonymous referee whose comments greatly improved the final
manuscript. 

\bibliographystyle{mn2e_w}
\bibliography{./mybib}
\end{document}